# Nuclear fusion enhances cancer cell killing efficacy in a protontherapy model

GAP Cirrone*, L Manti, D Margarone, L Giuffrida, A. Picciotto, G. Cuttone, G. Korn, V. Marchese, G. Milluzzo, G. Petringa, F. Perozziello, F. Romano, V. Scuderi

* Corresponding author

**Abstract**

Protontherapy is hadrontherapy's fastest-growing modality and a pillar in the battle against cancer. Hadrontherapy's superiority lies in its inverted depth-dose profile, hence tumour-confined irradiation. Protons, however, lack distinct radiobiological advantages over photons or electrons. Higher LET (Linear Energy Transfer) $^{12}$C-ions can overcome cancer radioresistance: DNA lesion complexity increases with LET, resulting in efficient cell killing, i.e. higher Relative Biological Effectiveness (RBE). However, economic and radiobiological issues hamper $^{12}$C-ion clinical amenability. Thus, enhancing proton RBE is desirable. To this end, we exploited the p + $^{11}$B→3α reaction to generate high-LET alpha particles with a clinical proton beam. To maximize the reaction rate, we used sodium borocaptate (BSH) with natural boron content. Boron-Neutron Capture Therapy (BNCT) uses $^{10}$B-enriched BSH for neutron irradiation-triggered alpha-particles. We recorded significantly increased cellular lethality and chromosome aberration complexity. A strategy combining protontherapy's ballistic precision with the higher RBE promised by BNCT and $^{12}$C-ion therapy is thus demonstrated.

The urgent need for radical radiotherapy research to achieve improved tumour control in the context of reducing the risk of normal tissue toxicity and late-occurring sequelae, has driven the fast-growing development of cancer treatment by accelerated beams of charged particles (hadrontherapy) in recent decades (1). This appears to be particularly true for protontherapy, which has emerged as the most-rapidly expanding hadrontherapy approach, totalling over 100,000 patients treated thus far worldwide (2). The use of energetic protons for cancer radiotherapy was first proposed by Wilson in 1946 (3). The primary motivation for investigation into this area was based on the physical properties of charged particles, which can deposit energy far more selectively than photons: through the inverted depth-dose profile described by the Bragg curve (4), healthy tissues within the entry channel of the beam are spared of dose, while most of the dose is steeply confined at the end of the particle range (the so-called "Bragg peak"). This in principle enables the delivery of very high-dose gradients close to organs at risk, confining the high-dose area to the tumour volume. Despite the dearth of randomized trials showing an effective advantage of protons over photon-based radiotherapy (5,6) and the ongoing debate over its cost-effectiveness (7), the current phase I/II clinical results support the rationale of the approach, especially for deep-seated tumours localized in proximity of critical organs, and unresectable or recurrent tumours (8,9). Cancer treatment by protons also remains the most attractive solution in the case of paediatric patients due to the significant reduction in the integral dose delivered to the patient (8), even compared to newer photon techniques such as intensity modulated radiation therapy (10). However, protons have been traditionally regarded as only slightly more biologically effective than photons (11). In fact, the standard practice in protontherapy is to adopt an RBE (Relative Biological Effectiveness) value of 1.1 compared to photons in any clinical condition (12), although such an assumption overlooks the increased RBE of low-energy protons (13-15) disregarding recently unveiled peculiarities of proton radiobiology (8,16,17).

The combination of ballistic precision with an increased ability to kill cells is the radiobiological rationale currently supporting the clinical exploitation of heavier particles such as fully stripped

$^{12}$C-ions (19), which present some advantages over protons (6, 19). Not only do they ensure a better physical dose distribution, due to less lateral scattering (20), but they are also more biologically effective both *in vitro* and *in vivo* as a result of their higher Linear Energy Transfer (LET) (11,21,22). In fact, densely ionizing radiation tracks cause more spatio-temporally contiguous and complex lesions at the DNA level, comprising DNA double-strand breaks and damaged bases, which are highly clustered in nature (23-25). This impairs cellular ability for correct repair (26) and decreases the dependence of radiosensitization upon the presence of oxygen, desirable features for eradication of resilient, hypoxic tumors (5, 27). Further potential radiobiological advantages include greater RBE for killing putatively radioresistant cancer stem cells (28) and counteracting cancer invasiveness (29, 30), albeit the latter remains controversial (31). Finally, low doses of high-LET radiation appear to elicit stronger immunological responses compared to low-LET radiation (17).

On the other hand, nuclear fragmentation reactions incurred by the primary beam partially spoil the $^{12}$C superior dose distribution, depositing unwanted dose behind the target volume, as opposed to the sharp dose fall-off that characterizes protons (32). Moreover, there exists a substantial lack of understanding of the impact on normal cells of the exposure to high-LET radiation. While it is true that preliminary reports on the follow-up of patients treated with $^{12}$C ions do not show an increased incidence of secondary cancers (33), uncertainties surround the effectiveness with which high-LET radiations such as $^{12}$C ions cause non-cancer effects that can have clinical repercussions. As a matter of fact, mounting evidence seems to indicate that $^{12}$C ions are particularly effective at causing pro-inflammatory responses leading to cardiovascular complications and premature cellular senescence even at low doses, i.e. those absorbed by the traversed healthy tissues (34-36). Taken together, these findings warrant further radiobiological investigation on the possible after effects arising from $^{12}$C hadrontherapy. Finally, a major hindrance to a greater diffusion of $^{12}$C hadrontherapy, which remains limited to a few centres, is caused by the complexity of their acceleration, transport and handling, delivery to the patient and dosimetry. These are more complicated compared to proton beams, leading to an increase of the initial investment and overall running costs of $^{12}$C-ion therapy centers (37).

In order to overcome such limitations, different strategies are being explored with the aim to achieve alternative solutions for a localized increase of proton RBE. One of the recently proposed approaches foresees the use of gold nanoparticles as protontherapy radiosensitizers (38). The ability of particle radiation to stimulate favourable immunological responses represents another attractive solution as it has become increasingly evident that proton and photon irradiation differentially modulate systemic biological responses (8, 17).

In this work, we experimentally demonstrate for the first time that the p + $^{11}$B $\rightarrow$ 3$\alpha$ nuclear fusion reaction, which is known to generate short-range high-LET alpha particles (39, 40), can be exploited to enhance proton biological effectiveness exclusively in the tumour region, and thus is of potential clinical worth. Cells were irradiated with a clinical proton beam in the presence of sodium borocaptate (NA$_2$B$_{12}$H$_{11}$SH or "BSH"), which is a common agent used in BNCT) in its $^{10}$B-enriched form to selectively deliver boron in cancer cells (41). BNCT requires thermal neutrons to trigger the reaction where a single alpha particle with maximum energy of around 2.7 MeV and range in tissue of less than 10 μm is produced (42). In order to maximize the p + $^{11}$B $\rightarrow$ 3$\alpha$ fusion rate, we used the compound with natural occurring boron isotopic abundance (i.e. 80% $^{11}$B and 20% $^{10}$B). We observed a significant increase in proton-induced cytogenetic effects, both in terms of cell death, assessed as loss of proliferative potential by the clonogenic cell survival, and of induction of DNA damage. The latter was studied by chromosome aberrations revealed by Fluorescence-in Situ Hybridization (FISH) painting. Specifically, the markedly higher frequency of complex-type chromosome exchanges (a typical cytogenetic signature of high-LET ionizing radiation, see ref. 43 for example), which was found among boron-treated cells compared to proton-irradiated cells in the absence of BSH, points to the alpha particles generated in the nuclear fusion reaction as being responsible for the measured enhancement of proton biological effectiveness. These findings, therefore, yield important implications for cancer protontherapy.

## Results

**Experimental approach**

The proton-boron nuclear reaction considered in this work is usually formalized as p+ $^{11}$B —> 3α. It has a positive Q-value (8.7 MeV) and is often referred to as the "proton-boron fusion reaction", as the incident proton is completely absorbed by the $^{11}$B nucleus. This reaction has garnered interest since the 1930s (39, 40) because of the process' ability to produce copious numbers of alpha particles in an exothermic reaction.

According to the most recent studies and interpretations, p+ $^{11}$B —> 3α can be basically described as a two-step reaction (involving $^{12}$C and $^{8}$Be nuclei excitation) with a main resonance occurring at a 675 keV (center of mass energy) and where the maximum cross section of 1.2 barn is measured (44, 45). A more detailed description of the reaction is reported in *Methods*.

The emitted alpha particles exhibit a wide energy spectrum with a predominant energy around 4 MeV (46). Such a reaction has been considered very attractive for the generation of fusion energy without producing neutron-induced radioactivity (47,48).

Such a nuclear reaction may be even more useful as it could play a strategic role in medical applications improving the effectiveness of protontherapy. The potential clinical use of the p-$^{11}$B reaction has thus far only been theoretically investigated (49). Here we experimentally implement for the first time this innovative approach and actually measure the biological effects elicited by such a reaction.

Besides the advantage of using a neutron-free nuclear reaction, the value of this method is also based on the fact that the the p + $^{11}$B →3α cross section becomes significantly high at relatively low incident proton energy, i.e. around the Bragg peak region. As schematically depicted in Fig.1, a conventional proton beam used in protontherapy is drastically slowed down across the tumour, corresponding to the Bragg peak region. Thus, most of its energy (dose) is delivered to the tumour cells. Assuming a given concentration of $^{11}$B nuclei is present preferentially, but not exclusively, in the tumour, the presence of slow protons could trigger a series of fusion events generating alpha particles localized in the tumour region. In fact, most of the alpha particles generated in the proton-boron reaction have an average range in water of less than 30 μm, comparable with typical cell size. Hence, even if such particles are mainly produced outside the cell cytoplasm due to sub-optimal boron uptake, the probability that they would reach the nucleus and damage the DNA remains very high. Moreover, even if a non-negligible concentration of $^{11}$B nuclei is present in the healthy tissues surrounding the tumour, the number of fusion events (i.e. generated alpha particles), will be relatively low or completely absent due the non-favourable incident proton energy fluence away from the tumour region. This would lead to a more biologically effective particle dose localization, higher than that currently achievable with conventional protontherapy, thus to a more efficient treatment in terms of an increase in cancer cell lethality because of the clustered nature of the DNA damage caused by the high-LET alpha particles emitted in the tumour region. Hence, protontherapy could acquire the benefits of an enhanced RBE for cancer cell killing, similar to $^{12}$C ion hadrotherapy but without the above-mentioned complications of the latter.

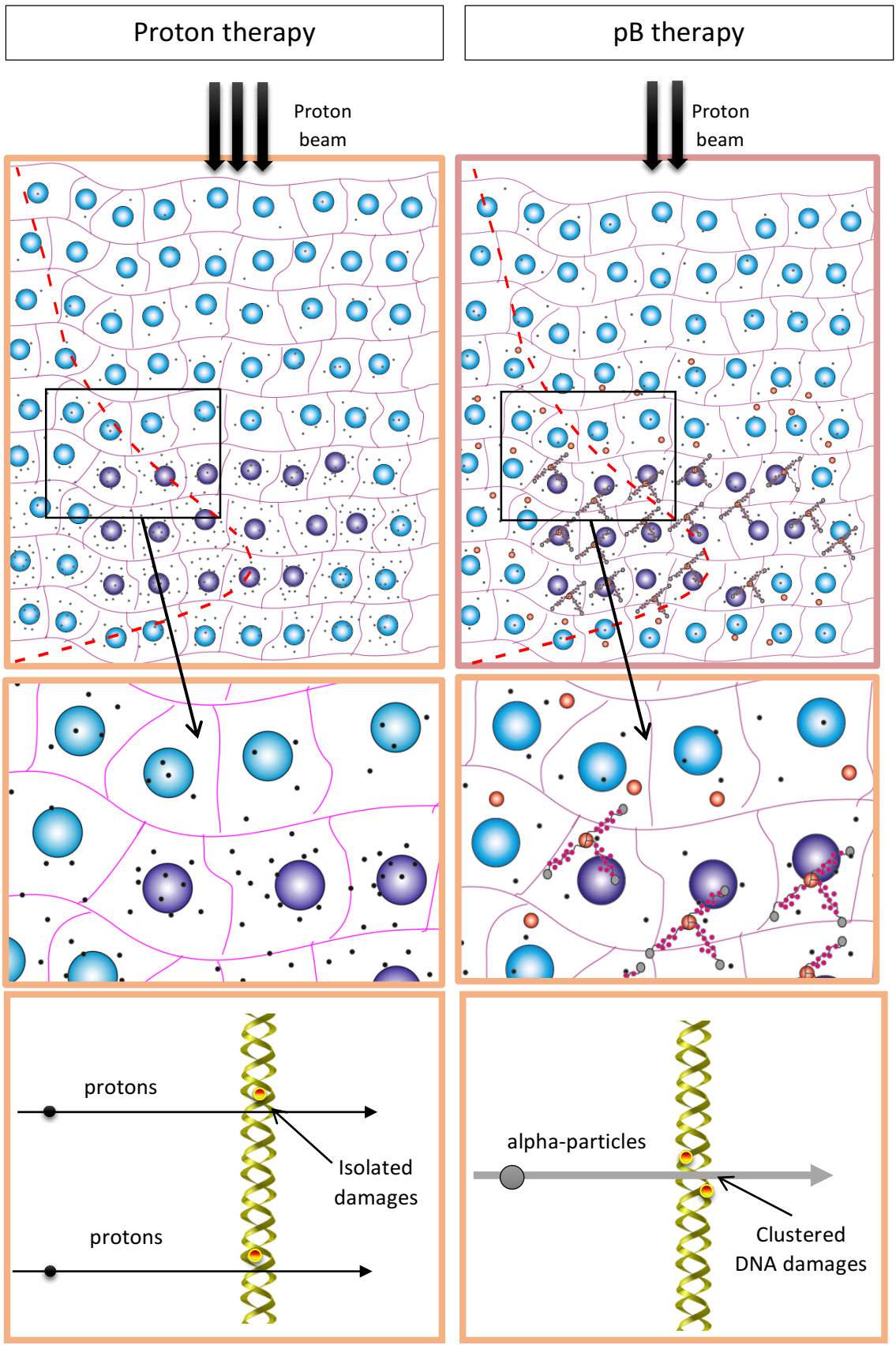

**Figure 1 | Schematic representation of "conventional" radiotherapy by low-LET proton beams (A) and the rationale of boron-enhanced protontherapy (B).**

The dose in protontherapy is currently released mainly in the tumor region (upper panel), cancer cells being represented here by purple circles and damaging events by black dots (**A**): proton-induced damage is similar to that imparted to DNA by photons, consisting mainly of isolated lesions (middle and lower panels). If cancer cells are loaded with $^{11}$B-delivering agents (middle panel, **B**), as is the case with $^{10}$B-enriched compounds in BNCT, unrepairable DNA clustered lesions will be also produced by the high-LET alpha particles generated by the p-$^{11}$B nuclear fusion reaction (lower panel). This in turn leads to the achievement of a greater RBE for cancer cell killing while maintaining beneficial sparing of surrounding healthy tissues (middle panel).

**BSH enhances cancer cell death following proton-irradiation**

To test whether the p + $^{11}$B →3α reaction results in an enhancement of cell killing by therapeutic proton beam irradiation, cells from the human prostate cancer line DU145 were irradiated with graded doses of the 60-MeV clinical proton beam available at the superconducting cyclotron of the INFN-LNS facility (Catania, Italy). Irradiations were performed in the presence of two concentrations of BSH (see *Methods* for details on the irradiation set-up and BSH pre-treatment). As a control, prostate cancer DU145 cells grown and irradiated without BSH were used. The considered BSH concentrations were equivalent to 40 ppm (parts per million) and 80 ppm of $^{11}$B. These were chosen based on values from the literature on the $^{10}$B-enriched BSH analogue used in BNCT in order to achieve the optimal $^{10}$B concentration (41, 50, 51). In particular, similar boron-equivalent concentration ranges of another BNCT compound, BPA, had been previously used with the same cell line *in vitro* (52). Boron treatment enhanced proton irradiation biological effectiveness resulting in a significant increase in the induction of cell death in DU145 cells as measured by loss of colony-forming ability. Cells that were irradiated after pre-treatment with, and in the presence of, boron-containing BSH exhibited a greater radiosensitivity in comparison with cells exposed to radiation alone: BSH-treated cells yielded a much steeper clonogenic dose-response curve than that obtained for cells grown and irradiated in BSH-free medium (Fig. 2). The magnitude of radiation-induced cell death was not affected by boron concentration. In our hands, the clonogenic survival fraction SF was best fitted to a linear function of dose, i.e. SF = exp (-α*D). Measured survival curve parameters were α = (0.79 ± 0.04) Gy$^{-1}$ for 80 ppm of $^{11}$B and α = (0.36 ± 0.06) Gy$^{-1}$ for the dose response curve obtained in the absence of BSH, with a calculated RBE$_{10}$ of 1.8. This indicated that protons in the presence of the boron compound were almost twice as effective at reducing cell survival by 90% compared to proton irradiation alone. Cellular radiosensitivity following low-LET proton irradiation was identical to that recorded after X-rays (Fig.2). At the concentrations used, BSH was not cytotoxic since the proliferative potential of unirradiated cells as given by cellular plating efficiency was not affected by the presence of BSH (Table 1). This means that the measured enhancement of proton effectiveness at cell killing was not contributed to by cytotoxicity caused by the boron-containing compound *per se*.

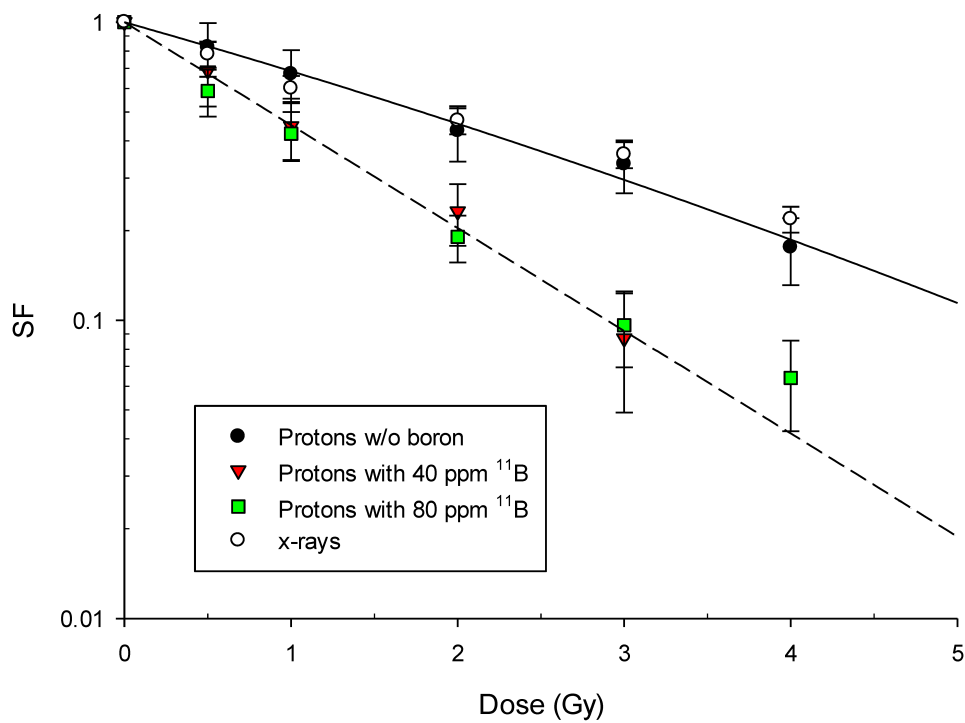

**Figure 2 | Boron-mediated increase in proton irradiation-induced cell death**. Clonogenic dose response curves of prostate cancer cells DU145 irradiated with therapeutic protons in the presence

or absence of BSH. X ray-irradiation survival data are also shown for comparison. For the sake of clarity, shown curves refer to least-square fitting of data from proton irradiation in the absence of BSH (closed circles) and in the presence of the compound at the highest concentration used (80 ppm, green squares). Data are weighted mean values plus standard error from at least two independent experiments.

|  | Plating efficiency | Baseline CA frequency |
|---|---|---|
| No BSH | 0.68 ± 0.13 | 0.024 ± 0.004 |
| 40 ppm $^{11}$B | 0.64 ± 0.17 | 0.023 ± 0.004 |
| 80 ppm $^{11}$B | 0.69 ± 0.11 | 0.018 ± 0.004 |

**Table 1**: **Cytogenotoxicity of BSH alone**. Plating efficiency (PE) values and total chromosome aberration (CA) yields in unirradiated DU145 prostate cancer cells and normal epithelial MCF-10A cells, respectively, as a function of the amount of BSH. By definition, PE measures the survival of cells in the absence of radiation. Similarly, the recorded frequency of CAs in cells not exposed to radiation is referred to as baseline CA frequency.

**Induction and complexity of proton irradiation-induced DNA damage are exacerbated by BSH**

Radiation-induced structural chromosome aberrations (CAs) arise from mis- or un-rejoined DNA breaks due to erroneous repair (53, 54). Ionizing radiation can give rise to a wide spectrum of aberration types (55). Complex-type exchanges, defined as those rearrangements involving at least two chromosomes and generated by at least three breaks, are an archetypical feature of high-LET exposure (43, 56). Measurement of CA frequency and, in particular, quantification of the proportion of complex-type chromosome exchanges was therefore instrumental to assess a possible increase in radiation-induced genotoxicity and whether this could be ascribed to high-LET alpha particles generated by the BSH-assisted p + $^{11}$B → 3α nuclear reaction. Cancer cells are known to be genetically unstable, hence they do not lend themselves to reliable assessment of radiation-induced DNA damage. Radiation-induced chromosome rearrangements would superimpose onto an elevated confounding frequency of baseline damage. Therefore, the non-tumorigenic breast epithelial MCF-10A cell line was chosen for scoring of radiation induced CAs. Proton irradiation resulted in a higher CA frequency in cells treated with BSH compared to cells irradiated with protons in the absence of BSH (Fig. 3). $^{11}$B concentration did not affect the measured frequency of radiation-induced CA as shown by the similar induction of DNA damage between 40 ppm and 80 ppm of $^{11}$B. In addition, in cells not exposed to radiation, BSH *per se* did not cause significantly higher genotoxic damage compared BSH-untreated cells (Table 1). The yield of CAs following X-rays was basically identical to that measured after exposure to protons in the absence of BSH (Fig. 3). Aberration data were fitted to a linear –quadratic function of the type $Y = Y_0 + \alpha * D + \beta * D^2$ where the coefficient $Y_0$ corresponded to the baseline CA frequency as reported in Table 1. Following irradiation in the absence of BSH, the values for the parameters were $\alpha = (0.003 \pm 0.021)$ Gy$^{-1}$ and $\beta = (0.024 \pm 0.007)$ Gy$^{-2}$, whereas α and β were $(0.05 \pm 0.03)$ Gy$^{-1}$ and $(0.021 \pm 0.007)$ Gy$^{-2}$ for proton irradiation in the presence of 80 ppm of $^{11}$B.

Because of the purely quadratic nature of the dose-response curve for proton irradiation in the absence of BSH (the α parameter was not statistically different from zero), no estimate for RBE$_{max}$ could be derived as this is defined as the ratio of the linear components of the linear-quadratic dose-response curve (57). RBE values were calculated for two levels of damage instead, that is 10 and 20 aberrations per 100 cells (58). In the case of 20 aberrations per 100 cell, the calculated RBE was about 1.4, whereas an RBE value of 1.7 was obtained for the level of 10 aberrations per 100 cells.

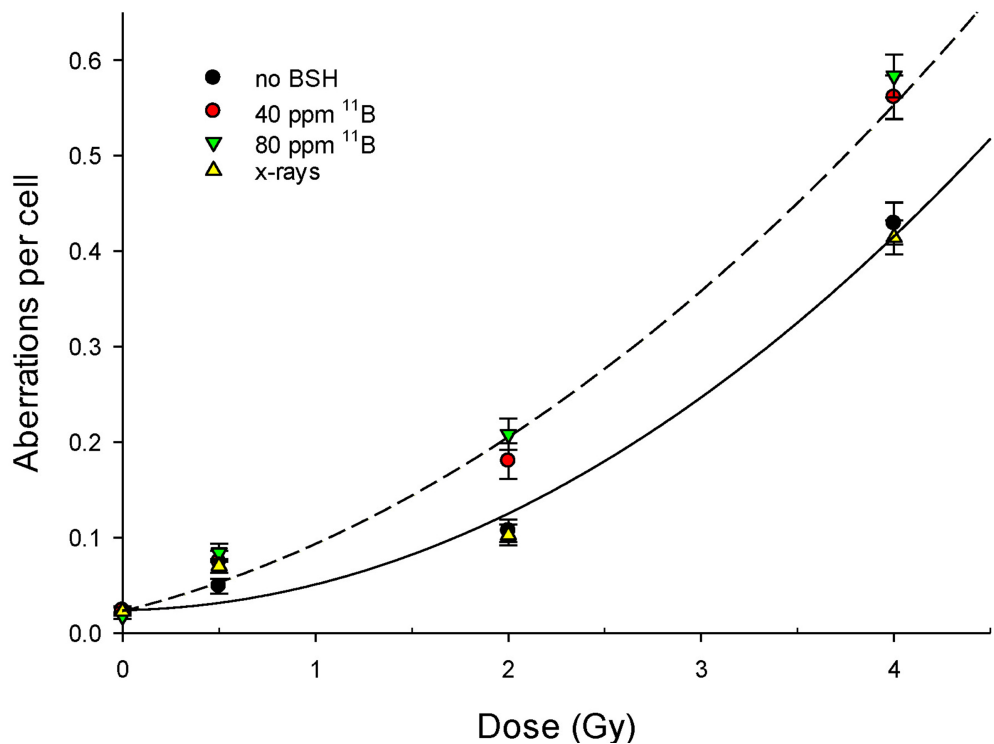

**Figure 3 | BSH-induced increased induction of chromosome aberrations following proton irradiation.** The dose-dependent frequency of all scored chromosome exchanges is shown for proton irradiation alone (black circles) and for proton irradiation in the presence of $^{11}$B at concentrations of 40 ppm (red circles) and 80 ppm (down triangles). X-ray data are also shown for comparison. In the interest of clarity, fitted curves are shown only for the highest boron concentration used (80 ppm, dashed line) and for irradiation with no boron compound (solid line). Data points correspond to the mean value measured in two independent experiments with standard errors of counts.

Interestingly, a markedly pronounced occurrence of complex-type exchanges was found in samples treated with BSH compared to cells that had been irradiated with protons in the absence of BSH (Fig. 4). After 0.5 Gy of protons the frequency of complex CAs ranged between 0.025 and 0.028 for BSH-treated cells as opposed to less than 0.004 in the case of cells irradiated in the absence of BSH. These values dramatically increased with dose and remained consistently higher in the case of BSH-treated cells, reaching about 0.18 and 0.08 after 4 Gy of protons in the presence or the absence of BSH, respectively. Occurrence of complex-type exchanges following X-rays is also shown for comparison (Fig. 4) and does not differ from that measured for proton irradiation alone.

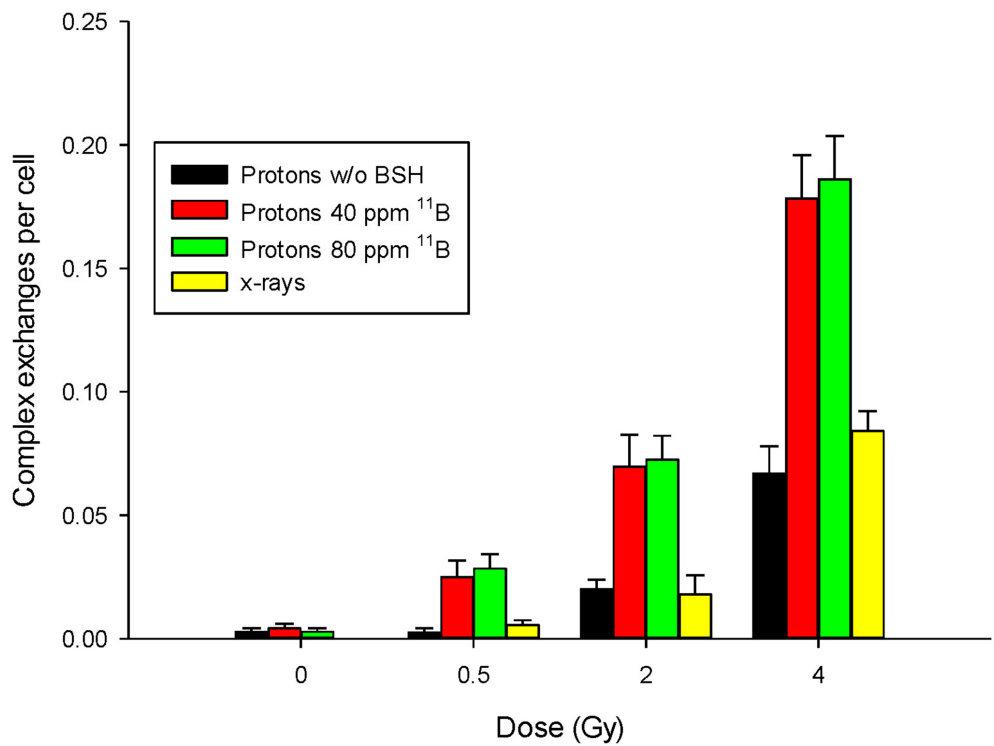

**Figure 4 | Induction of complex-type CAs**. A greater proportion of complex chromosome rearrangements was found in cells irradiated with protons and treated with BSH than in cells subject to proton irradiation alone.

Figure 5 shows pictures of a non-aberrant chromosome spread, with the two pairs of painted homolog chromosomes 1 and 2 clearly visible, and of a cell with chromosome rearrangements of complex nature as several portions of the painted chromosomes are aberrantly joined with aspecifically stained chromosomes (appearing blue) and with themselves. Taken together, these results strongly suggest high-LET radiation-induced damage, thus consistent with the action of the alpha particles produced by the proton-boron fusion reaction.

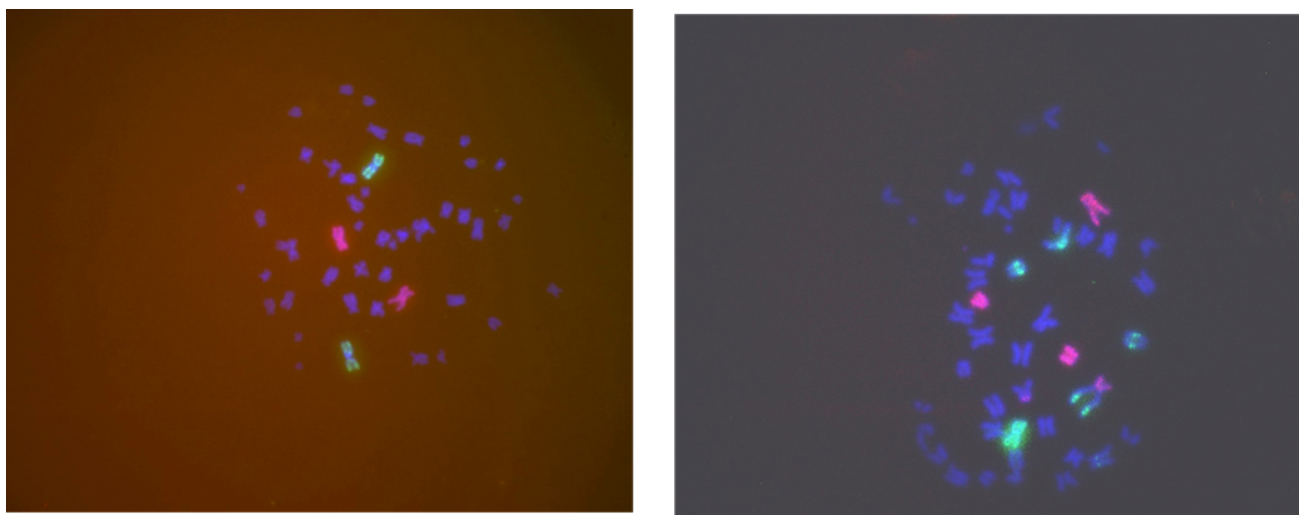

**Figure 5| Analysis of proton irradiation-induced structural chromosome damage.** Representative pictures of a non-aberrant normal chromosome spread (left) and of one from 4 Gy-proton irradiated-MCF-10A cells treated with 80 ppm of $^{11}$B exhibiting a complex-type CA (right).

## Discussion

For the first time, this work clearly proves that cellular irradiation with a clinical proton beam in the presence of a compound with natural boron isotopic content (80% of $^{11}$B + 20% of $^{10}$B) results in a significant enhancement of proton effectiveness at causing cytogenetic damage. The advantage of this new methodology is the fact that by using the considered fusion reaction, which has a high cross section for protons having energies up to 10 MeV, i.e. around the Bragg peak, the only products are three alpha particles with a range comparable to the cells' dimensions. The compound is BSH, one of the most commonly used boron-delivering agents in BNCT, where artificially $^{10}$B-enriched BSH must be used to produce alpha particles. The rationale underlying BNCT is that if cancer cells are selectively drugged with $^{10}$B, thermal neutron irradiation will result in a highly localized and mostly lethal targeting of cancer cells because of the very short range and high LET of the produced alpha particles. However, despite numerous and carefully designed clinical trials, BNCT has never quite fulfilled its potential as a powerful cancer treatment modality both because of the intrinsically challenging quest for ideal carriers to deliver radiobiologically effective concentrations of boron to cancer cells, and of the availability of thermal neutron sources (41, 51). On the other hand, albeit already being a clinical reality, external beam $^{12}$C-ion hadrontherapy, which is capable of delivering very biologically effective radiation doses with extremely high precision to the tumor target, is hampered by economic and radiobiological issues as illustrated in the *Introduction*.

Currently, the most widespread form of cancer treatment by accelerated charged particles is represented by protontherapy, which guarantees the same ballistic precision as $^{12}$C ion beams without the added complications. The disadvantage is a low biologically effectiveness since protons are only about as effective as photons and electrons at killing cells. The approach we propose and have successfully tested here could represent an enormous step towards the ability to increase the biological efficacy of proton radiotherapy by coupling the already favorable spatial and radiobiological characteristics of protons with the capacity to trigger the proposed reaction exclusively inside the tumor. In principle, this would enable the avoidance of the intrinsic uncertainties and enormous handling complications associated with the use of neutron beams in BNCT and could lead to an increase of proton biological effectiveness towards values closer to those exhibited by $^{12}$C-ions.

The biological effects of the proton-boron fusion reaction were investigated by measuring clonogenic cell death and chromosome aberrations (CAs) in a prostate cancer cell line (DU145) and in a non-tumorigenic epithelial breast cell line (MCF-10A), respectively. The latter was best suited to investigate chromosomal damage unlike genomically unstable cancer cells.

We found that proton irradiation-induced cellular lethality was greatly enhanced by irradiating cells that had been pre-treated with BSH. In particular, our results for survival of DU145 cells following low-LET proton irradiation alone are in line with those obtained at another protontherapy facility by Polf et al. (59), who studied the radiosensitizing effects of Au nanoparticles. These authors actually reported an enhancement of proton biological effectiveness of smaller magnitude than that found by us and aided by the p + $^{11}$B →3α reaction. Moreover, cellular survival values found by us following proton-triggered alpha irradiation from BSH are essentially identical to those found by Yasui et al. (52), exposing DU145 cells to neutrons in the presence of BPA, the other most common boron-delivering agent together with BSH.

Investigation of structural DNA damage in the form of CAs not only confirmed that the effectiveness of cellular proton irradiation is enhanced by the presence of boron but it also strongly suggests that such an enhancement could be explained by the action of high-LET radiation, e.g. alpha particles emitted by p + $^{11}$B →3α reaction. Firstly, a greater aberration frequency was found among BSH-treated cells. Following the highest dose of protons used (4 Gy) in the presence of 40 ppm of $^{11}$B, the observed frequencies of dicentrics (0.144 ±0.004) and rings (0.064±0.008) are consistent with those of 0.171 ±0.0175 and 0.029 ± 0.007, respectively, found by Schmid et al. (60) in human lymphocytes at the highest dose of thermal neutrons used by them (i.e. 0.248 Gy) and at a similar $^{11}$B-equivalent concentration (30 ppm). In addition, a significantly higher proportion of complex chromosome rearrangements in BSH-treated cells compared to controls following proton irradiation was found pointing to an LET-dependent effect since complex CAs are a well-known cytogenetic signature of exposure to high-LET radiation. In fact, it is well known that the greater biological effectiveness of densely ionizing radiation is a direct consequence of the physical pattern of energy deposition events along and around its tracks. Low-LET ionizing radiation such as x- or γ-rays mainly damage cells through short-lived bursts of free radicals (e.g. reactive oxygen species), generated by its interaction with the intracellular environment. This causes isolated lesions at the DNA level, the most detrimental of which for cell survival are double-strand breaks (DSBs). However, the much denser thread of ionization events specific to track-structured high-LET particle radiations, results in many closely spaced clusters of multiply DNA damaged sites, comprising DSBs together with single-strand breaks and damaged bases, which cause the cellular repair system to be error-prone. Hence, such lesions either are left unrepaired, which explains the greater efficiency of high-LET radiations at cell killing, or undergo misrepair. In the latter case, since single densely ionization tracks can simultaneously cause breaks in far apart stretches of DNA, i.e. belonging to separate chromosome domains, defective repair will lead to a higher frequency of DNA mis-rejoining events occurring between several chromosomes that after low-LET radiation, the so-called complex-type exchanges. In BSH-treated irradiated cells, the proportion of such complex-type aberrations relative to total exchanges ranged between 30% and 42% while that for proton irradiation alone was between 6% and 15%. Moreover, the ratio of complex- to simple-exchanges found by us ranged between 0.59 and 0.71, similar to the figure of 0.64 found by others for 0.5 Gy of direct alpha particle irradiation in first-division human lymphocytes scored with the same technique used by us, i.e. FISH painting (61).

We did not perform measurements of actual $^{11}B$ incorporation by cells but both drug pre-treatment times and concentrations were chosen according to available literature in BNCT studies showing the best conditions to achieve the ideal boron concentration in cells (20 μg per grams of tissue or $10^9$ atoms per cell). Indeed, although $^{10}B$-enriched BSH is known for its poor permeabilization through cell membranes, its use is facilitated by the higher boron content compared to the other widely used boron-delivering compound, BPA. However, the alpha particles emitted via the p + $^{11}B$ →3α reaction have average energies around 4 MeV corresponding to ranges in tissues of a few tens of microns, which ensures that severe cellular DNA damage is caused even if BSH is not incorporated in the cell but sits on its membrane. Moreover, the ballistic properties of the incident proton beam whereby proton energies drastically decrease in correspondence with the tumour volume, ensure negligible nuclear fusion events even in the worst-case scenario where the delivery agent containing $^{11}$B nuclei is also present in the healthy tissues surrounding the tumour. Furthermore, the presence of $^{10}$B in the proposed method allows to trigger different nuclear reactions generating prompt gamma-rays, which could be potentially used in a simultaneous treatment-and-diagnostics approach (49,62,63).

In conclusion, if further confirmed by both *in vitro* and pre-clinical investigations, our results represent an important breakthrough in cancer radiotherapy, particularly in the treatment of the disease by accelerated proton beams since it may significantly enhance its effectiveness without foreseeable patients' health complications and added financial costs.

# Methods

**Cell cultures**. Prostate cancer cell line DU145 and the spontaneously immortalized, nontransformed human mammary epithelial MCF-10A cells (kindly donated by prof. K. Prise, School of Medicine, Dentistry and Biomedical Sciences, QUB, UK) were grown in 25-cm2 (T25) standard tissue culture flasks, routinely subcultured and maintained at 37°C in a humidified atmosphere (95% air, 5% CO2). DU145 cells were used to assess possible enhancement by boron of proton-induced cancer cell death while chromosomal DNA damage was investigated in MCF-10A cells. DU145 cells were grown in RPMI medium supplemented with 10% fetal bovine serum and 1% antibiotics. Two media were instead needed for MCF-10A cells, one for optimal growth conditions and the other for resuspension and quenching of trypsin, as described in detail by Debnath et al. (64) Two days before irradiation DU145 and MCF-10A cells were seeded in T25 flasks at 105 and 6 105 cells/flask, respectively.

**BSH preparation and treatment**. A 1-g vial of sodium mercaptododecaborate or N-BSH, $Na_2[B_{12}H_{11}SH]$, FW 219.87, was purchased from KATCHEM Ltd. (Czech Republic). The working concentrations of 80 and 40 ppm of $^{11}B$ corresponded to 0.17 mg/ml and 0.08 mg/ml of BSH, respectively. BSH was decanted at the necessary amounts according to the total volume of the medium, in which the compound was thoroughly dissolved by simple agitation just prior to cell treatment. Boron cellular conditioning started 7 hrs prior to irradiation: the cell growth medium was aspirated and replaced with 5 ml of BSH-containing medium. Ordinary BSH-free growth medium was replaced into flasks that were used as controls. Just before irradiation, flasks were completely filled with the respective media (with or without BSH) to prevent cells from drying since flasks are irradiated standing vertically in front of the beam.

**p + $^{11}B$ → 3α nuclear fusion**

The p + $^{11}B$ → 3α nuclear fusion reaction at low energy can be basically described as a two-step reaction due to its behaviour at the three resonant energies (0.162 MeV, 0.675 MeV and 2.64 MeV). Firstly, a proton, interacting with $^{11}B$, induces the formation of a $^{12}C^*$ compound nucleus formed in the 2- or 3- excited state. If the $^{12}C^*$ nucleus is formed in its 2- state, it will decay to the first 2+ state of $^{8}Be$ emitting one alpha-particle with l = 3 (65). If the $^{12}C^*$ nucleus is formed in its 3- state, then the primary alpha particle can be emitted either with l = 1 from the decay to the first 2+ $^{8}Be$ excited state, or with l=3 from the decay to the 0+ 8Be ground state. In either case, the remaining $^{8}Be$ (2+ or 0+) nucleus immediately decays into two secondary alpha-particles with l' = 2. Alpha particles emitted in the first stage present a well-defined energy distribution and are commonly referred to as α0 and α1 if the $^8$Be 2+ or the 0+ states are populated, respectively. A few authors (65, 66) report that a very unlikely fourth channel, characterized by a maximum cross section of 10 μb in the 2.0 - 2.7 MeV energy range (66), can also be activated. In this case the $^{12}$C* directly breaks into three α particles skipping the intermediate $^8$Be stage, which show a continuous energy distribution.

Figure 5 reports some of the available experimental data (65,66) for the total production cross section of the p + $^{11}$B → 3α reaction as function of the mass centre energy and for the $\alpha_1$ channel. For low energies (0.1-5 MeV) the reaction cross sections become significantly high, thus maximising the alpha particle production around the proton Bragg peak region, an advantageous feature for the alternative protontherapy approach proposed in this work.

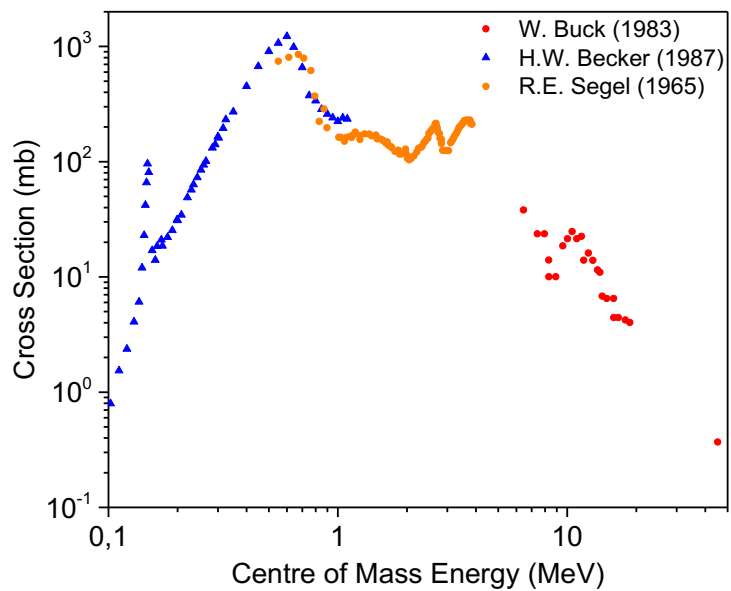

**Figure 5| Experimental cross sections for proton-$^{11}$B and the $\alpha_1$ channel**.

**Irradiation.** Irradiations were performed using the 62-MerV proton beam generated by the superconducting cyclotron clinically used at the CATANA (Centro di AdoTerapia ed Applicazioni Nucleari Avanzate) (67,68) eye proton therapy facility of the Italian Institute for Nuclear Physics in Catania, Italy. The CATANA irradiation setup for biological samples is described in detail elsewhere (69). The clinical Spread Out Bragg Peak (SOBP) range was 29 mm in water and cells were positioned at the depth of 18 mm water equivalent (calculated incident LET ~ 5 keV/mm), to the middle of the SOBP. Absolute absorbed dose is measured in water, by means of a plane-parallel PTW 34045 advanced-type Markus ionisation chamber, according to the International Atomic Energy Agency Technical Report Series 398 Code of practice. Overall uncertainty in absolute dose measurement is kept within 1.5%. Details on the irradiation beamlines, dosimetric procedures and related uncertainties for irradiation conditions can be found elsewhere (68). For measurement of cell killing, proton doses of 0, 0.5,1,2, 3 Gy were used. For 80 ppm of $^{11}$B the effects of a dose of 4 Gy were also tested. One cell culture flask was used for each dose for all BSH concentrations in each experiment. For chromosome aberration studies, three doses were used: 0.5, 2 and 4 Gy. Two flasks per dose were used for all BSH concentrations. Two independently repeated experiments were performed for both endpoints.

**Measurement of cell death.** Radiation-induced cell death was assessed by means of the clonogenic test, the gold standard for measuring cellular radiosensitivity. According to this assay, a cell survives irradiation if it retains its proliferative potential and it is thus capable of forming a colony composed of at least 50 cells. After irradiation, the medium was discarded from the flasks, and DU145 cells were detached, counted by haemocytometer and re-plated at opportune densities. They were then allowed to grow for colony formation in BSH-free medium. Four replicates for each dose were used for statistical data robustness. Clones were fixed and stained by crystal violet after 10 days. Surviving fractions (SF) are obtained by dividing the number of clones by the number of cells seeded at a given dose D, normalized by the plating efficiency (PE), that is the "surviving" of cells in the absence of radiation according to the expression $SF_{DoseD}$ = [Number of clones/number of cells]$_{DoseD}$/(PE). Experimental data were initially fitted to the linear-quadratic equation that best describes low-LET radiation induced cell death $SF = \exp(-\alpha*D-\beta*D^2)$. In all cases, however, the β parameter was found to be consistent with zero.

**Chromosomes aberration analysis.** After irradiation, MCF-10A cells were kept in BSH-free medium for up to 36 h. Genotoxic action of radiation was studied by scoring structural aberrations in chemically induced interphase chromosomes according to the Premature Chromosome Condensation (PCC) technique. Cells were incubated for 30 min with calyculin A (50 ng/mL, Sigma-Aldrich) for PCC induction as elsewhere described (70). To collect PCC spreads, cells were trypsinized, centrifuged at 300g for 8 min, then the pellet was resuspended for 25 min in hypotonic solution (75 mM KCl at 37°C), and fixed on ice in freshly prepared Carnoy solution (3:1 v/v methanol/acetic acid). Spreads were then dropped on pre-warmed (42 °C) wet slides and air-dried at room temperature. Fluorescence *In Situ* Hybridization painting was conducted by using whole-chromosome fluorescence-labelled DNA painting probes directed to chromosomes 1 and 2 following the manufacturer's recommendations (MetaSystems, Germany). Denaturation (72°C for 3 min) followed by hybridization (37°C overnight) of slides was performed using a hand-free HYBrite chamber system. Unlabeled chromosomes were counterstained with 12 ml 4,6-diamidino-2-phenylindole (DAPI) staining. Slides were viewed with an epi-fluorescence microscope connected to a computer-controlled system (Metafer 4 software, MetaSystems,) for automated slide scanning and three-color image acquisition. Chromosome analysis was carried out on stored images. Scoring was conducted blind by the same scorer. Not less than 500 chromosome spreads for each dose were scored (at least 1,500 for nonirradiated controls). All types of aberrations were scored separately and categorized as simple exchanges (i.e. translocations and dicentrics), either complete or incomplete, acentric fragments and complex exchanges. For this study's purpose, however, data herein presented are relative to the total chromosomal exchange frequency and to the complex-type exchange frequency. Frequency of total aberration exchanges was fitted to the equation $Y = Y_0 + \alpha*D + \beta D^2$.

**Statistical analysis**

For analysis of the dose – response relationships for cell killing and chromosome aberration frequency, curve fitting was performed by nonlinear least square minimization (Marquardt's algorithm) using SigmaPlot 12.5 software (SPSS, USA). Poisson statistics was assumed to derive standard errors on aberration frequencies. Experimental surviving fraction data in clonogenic assays are affected by several sources of errors, such as those associated with cell counts and cell dilutions, which are not taken into account by simple calculations of the standard error affecting colony counting. A more precise approach is needed to determine the experimental error on the plating efficiency PE as above defined and here recalled for convenience:

$$PE = \left(\frac{\bar{X}_{colonies}}{\bar{X}_{cells}}\right)\bigg|\ 0\ Gy$$

This quantity is the ratio between two mean values, i.e. the mean counted colony number and the mean number of cells seeded, each with its standard error SE: $SE_{colonies}$ and $SE_{cells}$. Therefore, according to the standard formula on error propagation, the standard error for plating efficiency, i.e. SE (PE), can be derived as follows:

$$SE(PE) = \sqrt{\left(\frac{\partial PE}{\partial \bar{X}_{colonies}}\right)^2 (SE_{colonies})^2 + \left(\frac{\partial PE}{\partial \bar{X}_{cells}}\right)^2 (SE_{cells})^2} =$$

$$\sqrt{\left(\frac{1}{\bar{X}_{cells}}\right)^2 (SE_{colonies})^2 + \left(-\frac{\bar{X}_{colonies}}{\bar{X}_{cells}^2}\right)^2 (SE_{cells})^2}$$

Recalling the definition of Surviving Fraction at a given dose D:

$$SF_D = \left(\frac{\bar{X}_{colonies}}{\bar{X}_{cells}}\right)\bigg/ PE$$

In the interest of simplicity, it can be assumed that the mean number of cells seeded that appears in the above formula is devoid of error, which is is reasonable since cell counting error is taken care of in the SE (PE). This means treating $\bar{X}_{cells}$ as a constant; hence we can define $\frac{\bar{X}_{colonies}}{\bar{X}_{cells}}$ as SF, with its error being $(SE_{colonies})/(\bar{X}_{cells})$

Hence, $SE(SF_D)$ can be determined as follows:

$$SE(SF_D) = \sqrt{\left(\frac{1}{PE}\right)^2 \left(\frac{SE_{colonies}}{\bar{X}_{cells}}\right)^2 + \left(-\frac{SF}{(PE)^2}\right)^2 (SE(PE))^2}$$

In these calculations we assumed that observed CV is always greater than Poisson CV on all counts, of either colonies or cells; wherever this was no the case, we corrected the experimental SE by multiplying the mean (of cell or colony counts) by the Poisson CV.

## Acknowledgments

This research was sponsored by the INFN (Italian Institute for Nuclear Physics) and by the project ELI - Extreme Light Infrastructure – Phase 2 (CZ.02.1.01/0.0/0.0/15_008/0000162) through the European Regional Development Fund. The authors gratefully acknowledge Rachael Jack for the support provided to improve the readability of the manuscript and the Director and Accelerators technical staff of INFN-LNS for providing the extra beam for the experiment

## Author contribution

Dr GAP Cirrone contributed to the conceptual idea of dose enhancement, proposed the biological measure, organized the experimental measures and followed all the dosimetric procedures; he organized the manuscript preparation and finalization. Dr L. Manti designed and carried out the radiobiological experiment, analyzed data and contributed to the discussion of the results and the writing of the manuscript. Dr D. Margarone contributed to the conceptual idea of dose enhancement and prompt gamma ray generation, experiment participation and manuscript preparation. Dr L. Giuffrida contributed in the conceptual idea and calculation of dose enhancement and prompt gamma ray generation and figures preparation. Dr G Cuttone and G. Korn contributed to the conceptual idea of dose enhancement and to manuscript preparation. Dr. Picciotto supported the idea of fusion reaction for the described application, participated to the experiment and into the manuscript preparation. Dr G Petringa contributed to the review and study of the 11B(p,α) αα reaction and was in charge of the review, collection and plot of the reaction experimental cross sections. She also contributed to the preparation of the experimental set-up and to irradiation. Dr F.M. Perozziello contributed to the experimental work on cell survival and data analysis. Dr F Romano was in charge of the absolute dosimetry measurements, sample irradiation and contributed to the experiment preparation. Dr V Scuderi contributed to the review, study and interpretation of the 11B(p,a)aa reaction also evaluating the total alpha generated during the experiment. Dr V Marchese and Dr G Milluzzo contributed the sample irradiation, experimental set-up arrangement and dosimetric preparation.

## Additional information

## Competing financial interests

The main author, on behalf of whole collaboration, declare that there are NO competing financial interests related to this work.

## References

1. Nuclear Physics European Collaboration Committee (NuPECC), Nuclear Physics for Medicine, NuPECC Report, *European Science Foundation* (2014)
2. Particle Therapy Co-Operative Group. Particle Therapy Centers. Available at: http://www.ptcog.ch/ (*last accessed on line August 2016*).
3. Wilson, R.R. Radiological use of fast protons, *Radiology* **47**, 487-491 (1946).
4. Bragg, W. & Kleemann, R. On the α-particles of radium and their loss of range in passing through various atoms and molecules, *Phil. Mag*. **10**, 318-340 (1905).
5. Loeffler J.S. & Durante, M. Charged particle therapy--optimization, challenges and future directions. *Nat. Rev. Clin. Oncol*. **10**, 411-424 (2013).
6. Uhl, M., Herfarth, K. & Debus, J. Comparing the use of protons and carbon ions for treatment. *Cancer J*. **20**, 433-439 (2014).
7. Verma, V., Shah, C., Rwigema, J.M., Solberg, T., Zhu, X. & Simone, C.B. Cost-comparativeness of proton versus photon therapy. *Chin. Clin Oncol.* **doi: 10.21037/cco.2016.06.03.** (2016). [Epub ahead of print]
8. Tommasino, F. & Durante, M. Proton radiobiology. *Cancers* **12**, 7353-7381 (2015).
9. Doyen, J., Falk, A.T., Floquet, V., Hérault, J. & Hannoun-Lévi, J.M. Proton beams in cancer treatments: Clinical outcomes and dosimetric comparisons with photon therapy. *Cancer Treat. Rev*. **43**, 104-112 (2016).
10. Hall, E.J. Intensity-modulated radiation therapy, protons, and the risk of second cancers. *Int. J. Radiat. Oncol. Biol. Phys*. **65**, 1-7 (2006).
11. Kraft, G. The radiobiological and physical basis of radiotherapy with protons and heavier ions. *Strahlenther. Onkol.* **166,** 10–13 (1990).
12. International Commission on Radiation Units and Measurements (ICRU). Prescribing, Recording, and Reporting Proton-Beam Therapy (Report 78), *J ICRU* **7**, Oxford University Press, Oxford (2007).
13. Belli, M. et al. Inactivation of human normal and tumour cells irradiated with low energy protons. *Int.J. Radiat. Biol*. **76**, 831-839 (2000).
14. Chaudhary, P. et al. Relative biological effectiveness variation along monoenergetic and modulated Bragg peaks of a 62-MeV therapeutic proton beam: a preclinical assessment. *Int. J. Radiat. Oncol. Biol. Phys*. **90**, 27-35 (2014).
15. Paganetti, H. Relative biological effectiveness (RBE) values for proton beam therapy. Variations as a function of biological endpoint, dose, and linear energy transfer. *Phys. Med. Biol*. **59**, R419-472 (2014).

16. Girdhani, S., Sachs, R. & Hlatky, L. Biological effects of proton radiation: What we know and don't know. *Radiat. Res*. **179**, 257–272 (2013).

17. Durante, M. New challenges in high-energy particle radiobiology. *Br. J. Radiol*. **87**, 20130626 (2014).

18. Combs, S. E. & Debus, J. Treatment with heavy charged particles: Systematic review of clinical data and current clinical (comparative) trials, *Acta Oncol*. **52**, 1272–1286 (2013)

19. Schulz-Ertner, D. & Tsujii, H. Particle radiation therapy using proton and heavier ion beams. *J. Clin. Oncol*. **25**, 953–964 (2007)

20. Amaldi, U. & Kraft, G. Radiotherapy with beams of carbon ions. *Rep. Prog. Phys*. **68**, 1861–1882 (2005).

21. Suzuki, M., Kase, Y., Yamaguchi, H., Kanai, T. & Ando K. Relative biological effectiveness for cell-killing effect on various human cell lines irradiated with heavy-ion medical accelerator in Chiba (HIMAC) carbon-ion beams. *Int. J. Radiat. Oncol. Biol. Phys*. **48,** 241-250 (2000)

22. Facoetti, A. et al. In vivo radiobiological assessment of the new clinical carbon ion beam at CNAO. *Radiat. Prot. Dosim.* **166**, 379–382 (2015).

23. Ward, J.F. The complexity of DNA damage: relevance to biological consequences. *Int. J. Radiat. Biol.* **66**, 427–432 (1994).

24. Goodhead, DT. Initial events in the cellular effects of ionizing radiations: clustered damage in DNA. *Int. J. Radiat. Biol*. **65**, 7–17 (1994).

25. Hada, M. & Sutherland, B.M. Spectrum of complex DNA damages depends on the incident radiation. *Radiat. Res*. **165**, 223–230 (2006).

26. Gustafsson, A.S., Hartman, T. & Stenerlöw, B. Formation and repair of clustered damaged DNA sites in high LET irradiated cells. *Int. J. Radiat. Biol*. **91**, 820-826 (2015)

27. Raju, M. R., Amols, H.I., Bain, E., Carpenter, S.G., Cox, R.A. & Robertson, J.B. A heavy particle comparative study. Part III: OER and RBE.*Br. J. Radiol.* **51**, 712-719 (1978).

28. Held, K. D. et al. Effects of Charged Particles on Human Tumor Cells. *Front. Oncol*. **12**, 6-23 (2016)

29. Ogata, T. et al. Carbon ion irradiation suppresses metastatic potential of human non-small cell lung cancer A549 cells through the phosphatidylinositol-3-kinase/Akt signaling pathway. *J. Radiat. Res*. **52**, 374-379 (2011)

30. Fujita, M. et al. Carbon-ion irradiation suppresses migration and invasiveness of human pancreatic carcinoma cells MIAPaCa-2 via Rac1 and RhoA Degradation. *Int. J. Radiat. Oncol.Biol. Phys*. **93**,173-180 (2015)

31. Fujita, M., Otsuka, Y., Imadome, K., Endo, S., Yamada, S. & Imai, T. Carbon-ion radiation enhances migration ability and invasiveness of the pancreatic cancer cell, PANC-1, in vitro. **103**, 677-683 (2012).

32. De Napoli, M. et al. Carbon fragmentation measurements and validation of the Geant4 nuclear reaction models for hadrontherapy. *Phys. Med. Biol*. **57**, 7651-7671 (2012).

33. Kamada, T. et al. Carbon ion radiotherapy in Japan: an assessment of 20 years of clinical experience. *Lancet Oncol*. **16,** e93-e100 (2015).

34. Yan, X. et al. Cardiovascular risks associated with low dose ionizing particle radiation. *PLoS One*. **9**, e110269 (2014)

35. Delp. M.D., Charvat, J.M., Limoli, C.L., Globus, R.K. & Ghosh, P. Apollo Lunar astronauts show higher cardiovascular disease mortality: possible deep space radiation effects on the vascular endothelium. *Sci Rep*. **6**, 29901 (2016).

36. Helm, A*.,* Lee, R., Durante, M. & Ritter, S. The influence of C-ions and X-rays on human umbilical vein endothelial cells. *Front. Oncol*. **6**, 5 (2016)

37. Nickoloff, J.A. Photon, light ion, and heavy ion cancer radiotherapy: paths from physics and biology to clinical practice. *Ann. Transl. Med*. **3**, 336 (2015).

38. Lin, Y., McMahon, S.J., Paganetti, H. & Schuemann, J. Biological modeling of gold nanoparticle enhancers radiotherapy for proton therapy. *Phys. Med. Biol*. **60**, 4149-4168 (2015).

39. Oliphant, M. & Rutheford, L. Experiments on the transmutation of elements by protons, Proc. R. Soc. Lond. A **141**, 259-272 (1933).

40. Dee, P.I. & Gilbert, C. W. The disintegration of Boron into three α-particles. *Proc. R. Soc. Lond. A* **154**, 279 (1936).

41. Barth, R.F. From the laboratory to the clinic: How translational studies in animals have lead to clinical advances in boron neutron capture therapy. *Appl. Radiat. Isot*. **106**, 22-28 (2015).

42. Atsushi, D. et al. Tumour-specific targeting of sodium borocaptate (BSH) to malignant glioma by transferrin-PEG liposomes: a modality for boron neutron capture therapy. *Neurooncol*. **87**, 287–294 (2008).

43. Anderson, R.M., Marsden, S.J., Wright. E.G., Kadhim. M.A., Goodhead. D.T. & Griffin, C.S. Complex chromosome aberrations in peripheral blood lymphocytes as a potential biomarker of exposure to high-LET alpha-particles. *Int. J. Radiat. Biol.* **76**, 31-42 (2000).

44. Spraker, M.C. et al. The $^{11}$B(p, α)$^{8}$Be →α+α and the $^{11}$B(α,α)$^{11}$B reactions at energies below 5.4 MeV. *J. Fusion Energ.* **31**, 357-367 (2012).

45. Sikora, M.H. & Weller, H.R. A. new evaluation of the $^{11}$B(p,α)αα reaction rates, *J. Fusion. Energ.* **35**, 538-543 (2016).

46. Stave S. et al. Understanding the $^{11}$B(p,α)αα reaction at the 0.675 MeV resonance. *Phys. Lett.* B **696**, 26-29 (2011).

47. N. Rostoker, M.W. Binderbauer, H.J.& Monkhorst, Colliding beam fusion reactor, *Science* **278**, 1419-1422 (1997).

48. A. Picciotto, et al. Boron-proton nuclear-fusion enhancement induced in boron-doped silicon targets by low-contrast pulsed laser, *Phys. Rev. X* **4**, 031030 (2014).

49. L. Giuffrida, D. Margarone, G.A.P. Cirrone, A. Picciotto, Prompt gamma ray diagnostics and enhanced hadron-therapy using neutron-free nuclear reactions, *arXiv:1608.06778*

50. Barth, R.F., Coderre, J.A., Vicente, M.G. & Blue, T.E. Boron neutron capture therapy of cancer: current status and future prospects. *Clin. Cancer Res.* **11**, 3987-4002 (2005).

51. Schwint, A.E.& Trivillin, V.A. 'Close-to-ideal' tumor boron targeting for boron neutron capture therapy is possible with 'less-than-ideal' boron carriers approved for use in humans. *Ther Deliv.* **6**, 269-272 (2015).

52. Yasui, L., Kroc, T., Gladden, S., Andorf, C., Bux, S. & Hosmane, N. Boron neutron capture in prostate cancer cells. *Appl. Radiat. Isot.* **70**, 6–12 (2012).

53. Iliakis, G. et al. Mechanisms of DNA double strand break repair and chromosome aberration formation. *Cytogenet. Genome Res.* **104**, 14–20 (2004).

54. Bailey, S. M. & J.S. Bedford, J.S. Studies on chromosome aberration induction: what can they tell us about DNA repair? *DNA Repair (Amst.)* **5**, 1171–1181 (2006).

55. Savage, J.R. & Tucker, J.D. Nomenclature systems for FISH-painted chromosome aberrations. *Mutat. Res.* **366**, 153-161 (1996).

56. Brenner, D.J. & Sachs, R. K. Chromosomal "fingerprints" of prior exposure to densely ionizing radiation. *Radiat. Res.* **140**, 134–142 (1994).

57. George, K., Durante, M., Willingham, V., Wu, H., Yang, T.C. & Cucinotta, F.A. Biological effectiveness of accelerated particles for the induction of chromosome damage measured in metaphase and interphase human lymphocytes. *Radiat. Res.* **160**, 425-435 (2003).

58. Wojcik, A. et al. Chromosomal aberrations in peripheral blood lymphocytes exposed to a mixed beam of low energy neutrons and gamma radiation. *J. Radiol. Prot*. **32**, 261–274 (2012).

59. Polf, J.C., Bronk,L. F., Driessen, W. H. P., Arap W., Pasqualini R. & and Gillin M. Enhanced relative biological effectiveness of proton radiotherapy in tumor cells with internalized gold nanoparticles. *Appl. Phys. Lett*. **98**, 193702 (2011).

60. Schmid, T. E., Canella, L., .Kudejova, P., Wagner, F. M., Roöhrmoser, A. & Schmid, E. The effectiveness of the high-LET radiations from the boron neutron capture [$^{10}$B(n,)$^{7}$Li] reaction determined for induction of chromosome aberrations and apoptosis in lymphocytes of human blood samples. *Radiat. Environ. Biophys*. **54**, 91–102 (2015).

61. Barquinero, J. F., Stephan G. & Schmid, E. Effect of americium-241 ✓-particles on the dose–response of chromosome aberrations in human lymphocytes analysed by fluorescence in situ hybridization. *Int. J. Radiat. Biol*. **80**, 155-164 (2004).

62. Do-Kun, Y., Joo-Young, J. & Tae S.S. Application of proton boron fusion reaction to radiation therapy: A Monte Carlo simulation study. *Appl. Phys. Lett*. **105**, 223507 (2014).

63. Shin, H.-B., Yoon, D.K., Jung, J.Y., Kim, M.S. & Suh, T.S. Prompt gamma ray imaging for verification of proton boron fusion therapy: A Monte Carlo study. *Phys.Med*. http://dx.doi.org/10.1016/j.ejmp.2016.05.053. (2016).

64. Debnath, J., Muthuswamy, S. K. & and Brugge, J.S. Morphogenesis and oncogenesis of MCF-10A mammary epithelial acini grown in three-dimensional basement membrane cultures. *Methods* **30**, 256–268 (2003).

65. Becker, H.W., Rolfs, C. & Trautvetter, H. P. Low-energy cross sections for $^{11}$B(p,3✓), *Z. Phys. A Atomic Nuclei* **327**, 341-355 (1987).

66. Segel, R.E., Hanna, S.S. & Allas, R.G. States in C$^{12}$ between 16.4 and 19.6 MeV. *Phys. Rev*. **139**, 818-830 (1965).

67. Cuttone G., Amato A., Bartolotta A., Brai M., Cirrone G.A.P., Giammò A., Lo Nigro S., Nicoletti G.A., Ott J., Privitera G., Raffaele L., Rallo M.L., Rapicavoli C., Reibaldi A., Rifuggiato D., Romeo N., Rovelli A., Sabilini M.G., Salamone V., Teri G., Tudisco F. Use of 62 MeV proton beam for medical applications at INFN-LNS: CATANA project. *Physica Medica* **17** 23-25 (2001).

68. Cirrone G.A.P., Lo Jacono P., Lo Nigro S., Mongelli V., Patti I.V., Privitera G., Raffaele L., Rifuggiato D., Sabini M.G., Salamone V., Spatola C., Valastro M.L. A 62 MeV proton beam for the treatment of ocular melanoma at Laboratori Nazionali del Sud-INFN. *IEEE Trans. Nucl. Sci.* **51** 860-865 (2004).

69. Manti, L. et al., Measurements of metaphase and interphase chromosome aberrations transmitted through early cell replication rounds in human lymphocytes exposed to low-LET protons and high-LET $^{12}$C ions. *Mutat. Res*. **596**, 151-165. (2006).

70. Durante, M., Furusawa, Y. & Gotoh, E. A simple method for simultaneous interphase-metaphase chromosome analysis in biodosimetry. *Int. J. Radiat. Biol*. **74**, 457- 462 (1998).